# Numerical investigation of tuning method for nonuniform disk-loaded waveguides

M.I. Ayzatsky[1], V.V.Mytrochenko

National Science Center Kharkov Institute of Physics and Technology (NSC KIPT), 610108, Kharkov, Ukraine

On the base of more general model we assessed accuracy of method that is widely used for post- tuning accelerator sections. Our consideration has shown that for wide range of disk loaded waveguide (DLW) parameters using simple "tuning" coefficients gives appropriate characteristics of DLWs. For injector sections, in which phase velocity and amplitude change, there are some deviation from the desired characteristics. It is necessary to evaluate the influence of additional phase shift on beam characteristics at the injector exit.

## 1 Introduction

In order to provide synchronism with the beam and electromagnetic field in the accelerating structure, the phase advance of each cell needs to be adjusted to its nominal value. This can be done after brazing by correcting machining deviations, assembly and brazing mismatching. This adjustment process is called tuning (post-tuning).

The tuning methods based on the non-resonant perturbation field distribution measurement [1,2,3,4,5,6,7] have been widely used for tuning travelling-wave structures, especially in tuning the constant-gradient ones. [8,9,10,11,12,13,14,15,16,17,18,19,20,21]. There are three approaches for post-tuning.

In [13] as a tuning parameter was chosen a standing-wave ratio (SWR). All regular cells were divided into some periods, and in each period there were three cells. The SWR was defined as the largest field peak divided by the smallest field peak.

In [19] iterative procedure was used to tune a prototype crab cavity for CLIC. The tuning of each cell was repeated until the electric field pattern was satisfactory. Subsequently the next cell towards the input was tuned. Occasionally, a cell already tuned had to be revised. After 26 steps the nominal phase advance per cell was achieved.

But the most widespread tuning method became one, in which the field distribution was considered to be a linear superposition of forward and backward waves in each cell [8,9,10,11,14,15,16,17,18,20,21]. The internal reflection of each cell was obtained by calculating the difference of the backward waves seen before and after that cell. The normalized voltage of cell was introduced for nonuniform structures with great field gradients [16]. But forward and backward waves were not strictly determined. Their amplitudes were introduced phenomenologically [8,9]

Development of the new coupling cavity model gave possibility to look on the properties of the nonuniform DLWs more deeply, especially on the base of methods that are used for tuning nonuniform DLWs [22,23,24,25]

## 2 Characterization of detuned cells

Let's consider a cylindrical nonuniform DLW (Fig.1). We will consider only axially symmetric fields with $E_z, E_r, H_\varphi$ components. Time dependence is $\exp(-i\omega t)$.

---

[1] M.I. Aizatskyi, N.I.Aizatsky; aizatsky@kipt.kharkov.ua

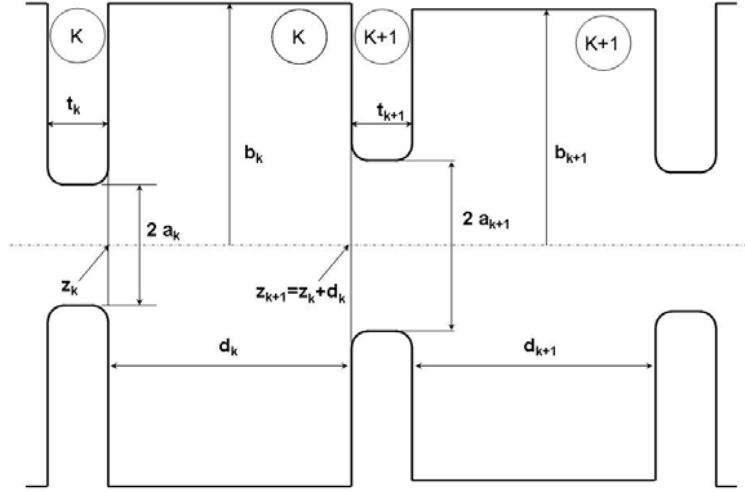

<div align="center">Fig. 1</div>

In the frame of Coupling Cavity Model (CCM) such coupling equations can be obtained [23,24]

$$Z_n e_{010}^{(n)} = \sum_{j=-\infty, j\neq n}^{\infty} e_{010}^{(j)} \alpha_{010}^{(n,j)} \qquad . \tag{1.1}$$

Here $e_{010}^{(n)}$ - amplitudes of $E_{010}$ modes, $Z_n = 1 - \dfrac{\omega^2}{\omega_{010}^{(n)2}} - \alpha_{010}^{(n,n)}$, $\omega_{010}^{(n)}$ - eigen frequencies of these modes, $\alpha_{010}^{(n,j)}$ - real coefficients that depend on the geometric sizes of coupling volumes. Sums in the right side can be truncated

$$Z_n^{(N)} e_{010}^{(N,n)} = \sum_{j=n-N, j\neq n}^{n+N} e_{010}^{(N,j)} \alpha_{010}^{(n,j)} \qquad . \tag{1.2}$$

For characterization of cell tuning we can use such coefficients (tuning parameters) [22]

$$F_n^{(2N+1,m)} = \frac{Z_n^{(N,m)}}{Z_n^{(N,c)}} - 1 . \tag{1.3}$$

where

$$Z_n^{(N,m)} = \frac{1}{\chi_n^{(N,c)} E_n^{(m)}} \sum_{j=n-N, j\neq n}^{n+N} \chi_j^{(N,c)} E_j^{(m)} \alpha_{010}^{(n,j)} , \tag{1.4}$$

$$\chi_n^{(N,c)} = \frac{e_{010}^{(N,n,c)}}{E_n^{(N,c)}} , \qquad . \tag{1.5}$$

$E_n^{(m)}$ - complex field amplitude measured in the middle of n-th cell, $e_{010}^{(N,n,c)}, E_n^{(N,c)}$ - calculated complex field amplitudes, $Z_k^{(N,c)}$, $\chi_n^{(N,c)}$ - calculated coefficients.

Note, that under certain values of $N$ and coupling strength (dependence of $\alpha_{010}^{(n,n)}$ on parameters of adjacent cells must be very small) the "tuning coefficients" (1.3) are local ones – they only depend on the parameters of the considered cell. This property is exactly what gives possibility to tune cells consecutively one by one.

After conducting full numerical simulation of the DLW and obtaining all necessary coupling coefficients (a new numerical method for DLW tuning [25] can be used for calculation the necessary values) someone can start the tuning process on the base of bead-pull field distribution measurements, that give the values of $E_n^{(m)}$. Making coefficients $F_n^{(2N+1,m)}$ for each cell as small as possible (zero in the limit) we tune the whole structure with some accuracy.

But there are some drawbacks of such approach. First one is the possibility conducting full numerical simulation of nonuniform DLW and obtaining all necessary coupling coefficients. The CCM, that was developed recently [22,23,24], can be used only for simple geometries for which there are analytical expressions for eigen functions. The second one is the accuracy of numerical simulation. This accuracy will be transferred on the tuning accuracy (together with the accuracy of measurements). For receiving the necessary accuracy of simulation we must take into account the great number of eigen functions that can be difficult from several reasons (accuracy of Bessel function calculation, etc). The CCM can be based on the general approach that gives possibility to obtain the coupling coefficients for arbitrary chain of resonators (see Appendix 1). But additional simulations are needed to carry out to obtain these coefficients.

There is case when the introduced tuning parameters take the form that practically do not depend on simulation results – case of small gradients of geometric parameters and small coupling of the adjacent cells. In this case we can choose $N = 1$ (cell "interacts" only with the adjacent cells) and consider that $\chi_n^{(1,c)} \approx \chi_{n\pm1}^{(1,c)}$, $\alpha_{010}^{(n,n+1)} \approx \alpha_{010}^{(n,n-1)}$, $Z_n^{(1,c)} \approx 2\alpha_{010}^{(n,n+1)} \cos\varphi$. Then we obtain

$$F_n^{(3,m)} \equiv F_n^{(m)} = \frac{Z_n^{(1,m)}}{Z_n^{(1,c)}} - 1 \approx \frac{E_{n-1}^{(m)} + E_{n+1}^{(m)} - 2\cos\varphi\, E_n^{(m)}}{E_n^{(m)}\, 2\cos\varphi}\,. \tag{1.6}$$

This approach was used by the most accelerator section manufacturers [8,9,10,11,14,15,16,17,18,20,21].

Accuracy of tuning on the base of these coefficients can be estimated only in the frame of more general model.

In this paper we present the results of comparisons of DLW characteristics of several pieces of nonuniform DLWs that were tuned with using the new numerical method [25] and tuning coefficients$(1.6)^2$. We will consider the DLWs with a constant phase shift per cell equals $\varphi = 2\pi/3$.

There is problem of taking into account absorption of RF energy in DLW walls as there are difficulties in obtaining appropriate eigen functions for cylindrical regions. All developed procedures in the frame of the CCM do not include this phenomenon. We used the simplest approach for including absorption into consideration. We supposed that the coupling coefficient do not depend on absorption and replaced resonant term in the equation for $E_{010}$ amplitudes (1.1) with

$$Z_n = 1 - \frac{\omega^2}{\omega_{010}^{(n)2}} - \alpha_{010}^{(n,n)} \Rightarrow 1 - i\frac{\omega}{\omega_{010}^{(n)}Q_n} - \frac{\omega^2}{\omega_{010}^{(n)2}} - \alpha_{010}^{(n,n)}\,, \,. \tag{1.7}$$

where $Q_n$ is a quality factor of the n-th cell.

## 3 Tuning of nonuniform disk-loaded waveguides

On the base of coupling equations (1.1) we can study the processes of wave propagation in inhomogeneous DLWs. It can be done if we suppose that before the inhomogeneous DLW there is a homogeneous fragment of DLW without absorption (an input waveguide) and after the inhomogeneous DLW there is also such fragment (an output waveguide). Two cells between the homogeneous parts and the inhomogeneous one are used for matching (couplers). In this case in homogeneous fragments of DLW at sufficient distance from the connection interfaces (when all evanescent wave decay) we can search amplitudes in the form

$$e_{010}^{(n)} = \begin{cases} \exp\{i\varphi_{1,0}(n-n_1+1)\} + R\exp\{-i\varphi_{1,0}(n-n_1+1)\},\ n < n_1, \\ T\exp\{i\varphi_{2,0}(n-n_2-1)\} \qquad\qquad\qquad\qquad n > n_2 \end{cases}, \tag{1.8}$$

where $R$ is the reflection coefficient and $T$ is the "transmission" coefficient.

---

[2] The first approach we will denote as NM tuning and the second one as TCM tuning

### 3.1 SLAC type DLW

One of the most popular structures is the SLAC constant gradient structure [26]. We consider the process of tuning of the front 11 cells of this structure with taking into account attenuation in the walls.

The input homogeneous DLW (cells from 1 to 14) consists of 14 cells with $a = 1.311$ cm ($d = 2.9148$ cm, $t = 0.5842$ cm, $r = .0$ cm, $Q = 1.E10$). The 15[th] cell is the input coupler which parameters ($a(15)$ and $b(15)$) were chosen under matching homogeneous DLWs with $Q = 1.E10$ and $Q = 1.2E4$ (Usually, the reflection coefficient $R$ was decreased to magnitude less than 1.E-4) .

The output homogeneous DLW (cells from 33 to 46) consists of 14 cells with $a = 1.2841$ cm ($d = 2.9148$ cm, $t = 0.5842$ cm, $r = .0$ cm, $Q = 1.E10$). The 32[th] cell is the output coupler which parameters ($a(33)$ and $b(32)$) were chosen after the fragment (cells N16-31) tuning by realizing the condition of minimum $R$ ( the reflection coefficient, see (1.8)).

The cells with N 18-28 are similar[3] to the front part of the SLAC structure.

The quality factors of cells from 15 to 32 ($a = 1.311$ cm, $d = 2.9148$ cm, $t = 0.5842$ cm, $r = .0$ cm) were chosen equals $Q = 1.2E4$[4].

The NM tuning procedure consisted of consecutive (from the beginning to the end) changes of cavity radii until the parameters $G_n^{(3)}$ (see [25]) for cell numbers N16÷29 differ from 1 greater than ±1.E-4. The reflection coefficient was achieved $R = 1.1E-4$.

The TCM tuning procedure consisted of consecutive (from the end to the beginning) changes of cavity radii until the parameters $F_n^{(m)}$ (see(1.6)) for cell numbers N16÷30 differ from 0 greater than ±1.E-4. The reflection coefficient was achieved $R = 9E-5$.

Dependences of radii of apertures on cell number after NM tuning[5] is presented in Fig. 2. Radii of cavities for two tuning procedures are presented in Fig. 3 and Fig. 4. We can see that differences of cavity radii for two tuning methods  for parameters of DLW under consideration are small and are less than ±2 µm.

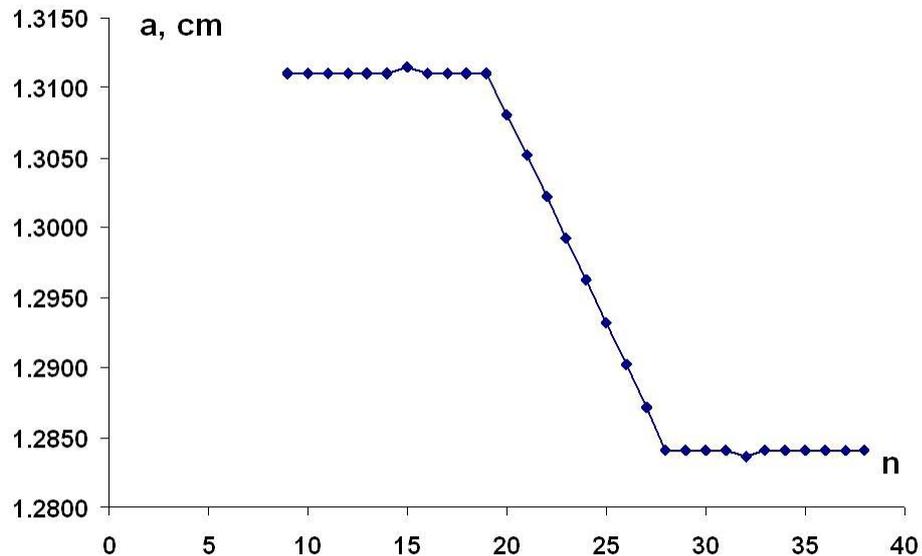

Fig. 2 Radii of disk apertures openings as a function of the cell number after TM tuning

---

[3] In real SLAC structure apertures have rounding.  We restrict ourselves by the case $r = 0$ to decrease  the simulation time

[4] This value slightly differs from quality factors of SLAC structure ($Q \sim 1.4E4$)

[5] These values were used for the TCM tuning too

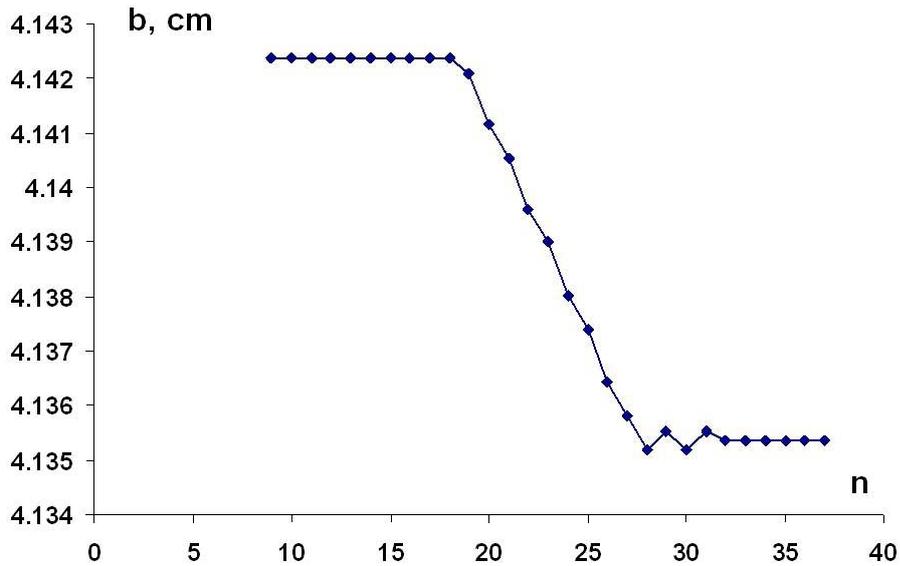

Fig. 3 Radii of cavities as a function of the cell number after TM tuning

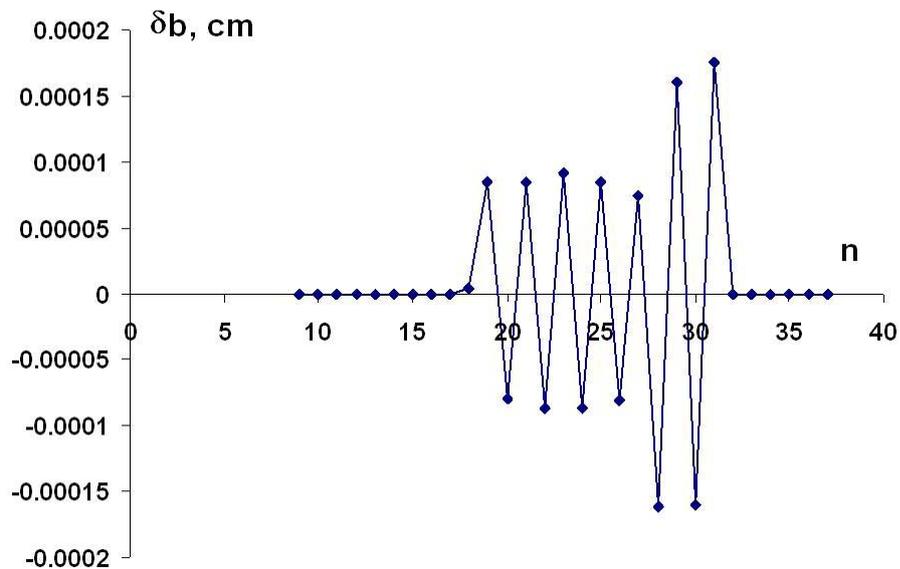

Fig. 4 Differences of cavity radius for two tuning methods (NM and TCM)

However, differences of values of $G_n^{(3)}$ and $F_n^{(m)}$ for NM and TCM tuning methods are rather large (Fig. 5, Fig. 6).

It is useful to draw attention to the imaginary part of $F_n^{(m)}$. This parameter has no physical sense in the frame of model with forward and backward waves [8]. Fig. 7 shows that in the part with nearly constant gradient (cells N 18-28) the imaginary part of $F_n^{(m)}$ is near zero. In cells with constant impedance where $Q = 1.\text{E}10$ $F_n^{(m)}$ (real and imaginare parts) equals zero and where $Q = 1.2\text{E}4$ the imaginary part of $F_n^{(m)}$ is relatively large. From this results we can conclude that the he imaginary part of $F_n^{(m)}$ proportional to the amplitude gradient. In the case of section with constant impedance the imaginary part of $F_n^{(m)}$ is proportional to the attenuation coefficient $(1/Q)$.

Characteristics of electromagnetic fields in the structure after tuning are presented in Fig. 8-Fig. 11. From these results it follows that the DLW have the similar electromagnetic field distributions with using for tuning two different methods[6]. The maximum phase deviation along three cells[7] is less than 0.5 degree.

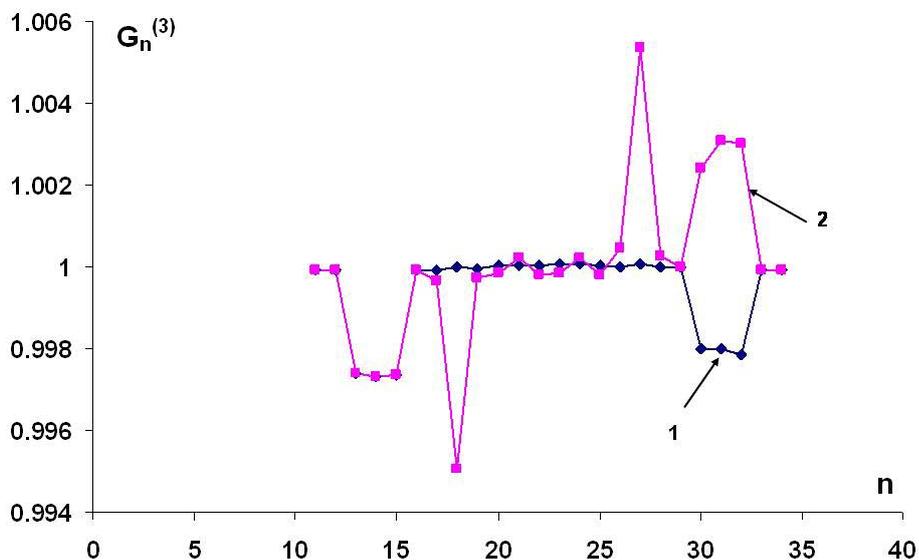

Fig. 5 $G_n^{(3)}$ for NM (1) and TCM (2) tuning methods

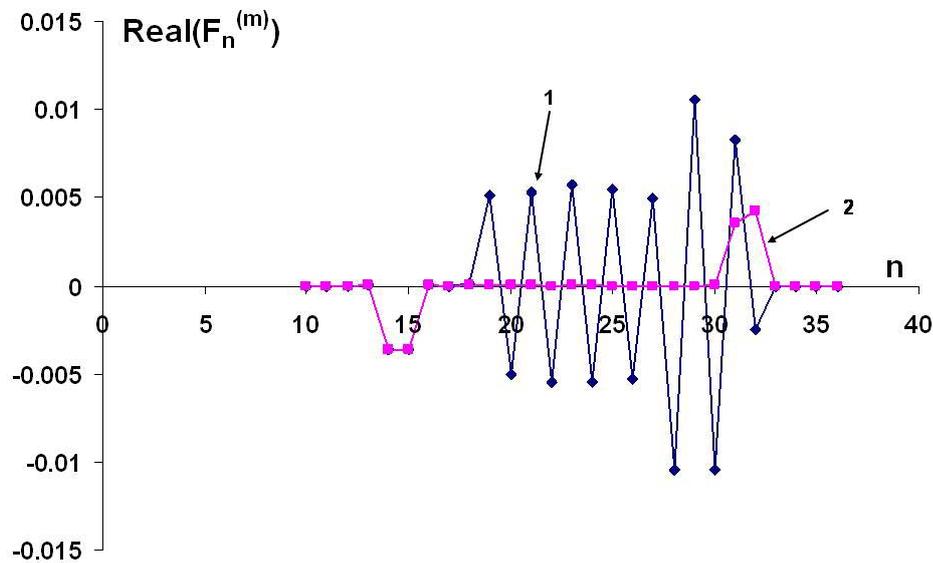

Fig. 6 Real part of $F_n^{(m)}$ for NM (1) and TCM (2) tuning methods

---

[6] We remind that the NM is purely the computer tuning method that is based on using the coupling coefficients. The TCM is based on the using values of longitudinal electric field in the middle of the cells. So, it can be realized on the base of measured or calculated values of longitudinal electric field.

[7] By choosing the proper initial phase from which a particle starts its movement

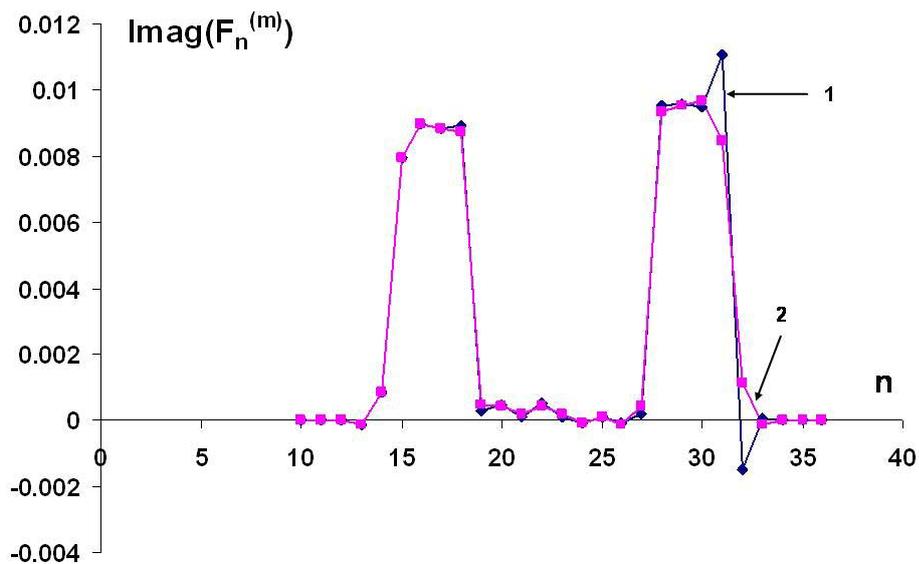

Fig. 7 Imaginary part of $F_n^{(m)}$ for NM (1) and TCM (2) tuning methods

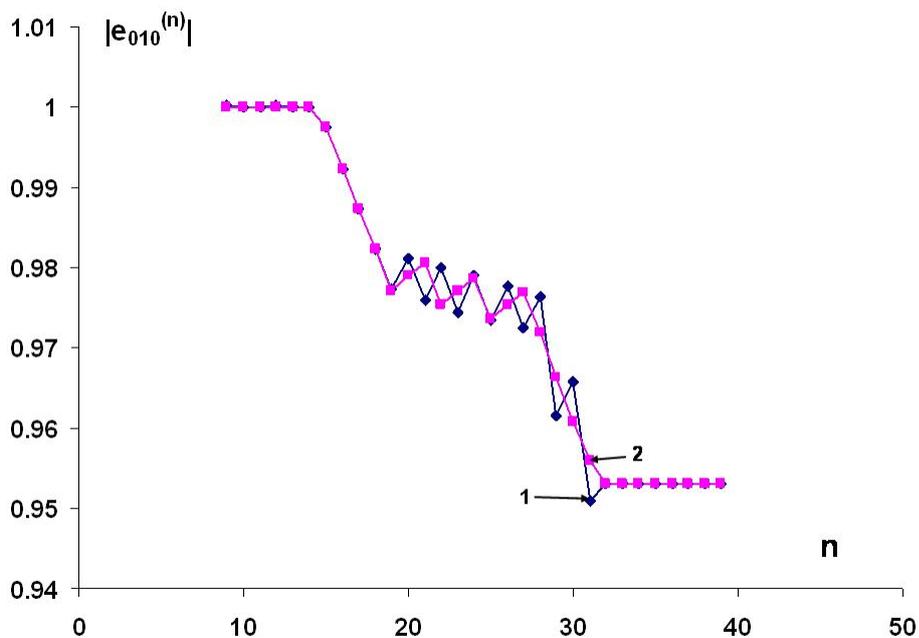

Fig. 8 Absolute values of amplitudes of $E_{010}$ eigen mode for NM (1) and TCM (2) tuning methods

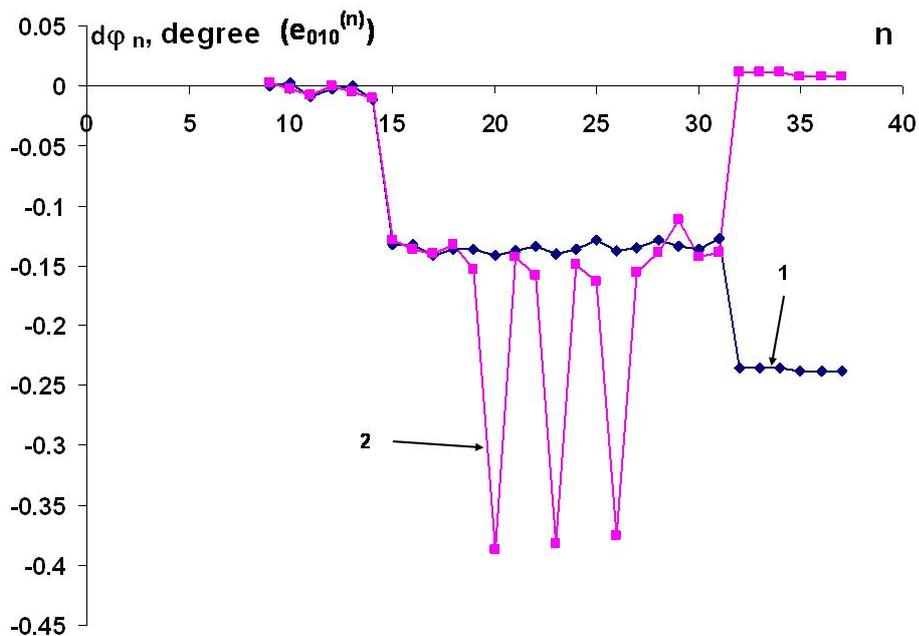

Fig. 9 Deviations of phase shift of amplitude of $E_{010}$ eigen mode from $2\pi n/3$ for NM (1) and TCM (2) tuning methods

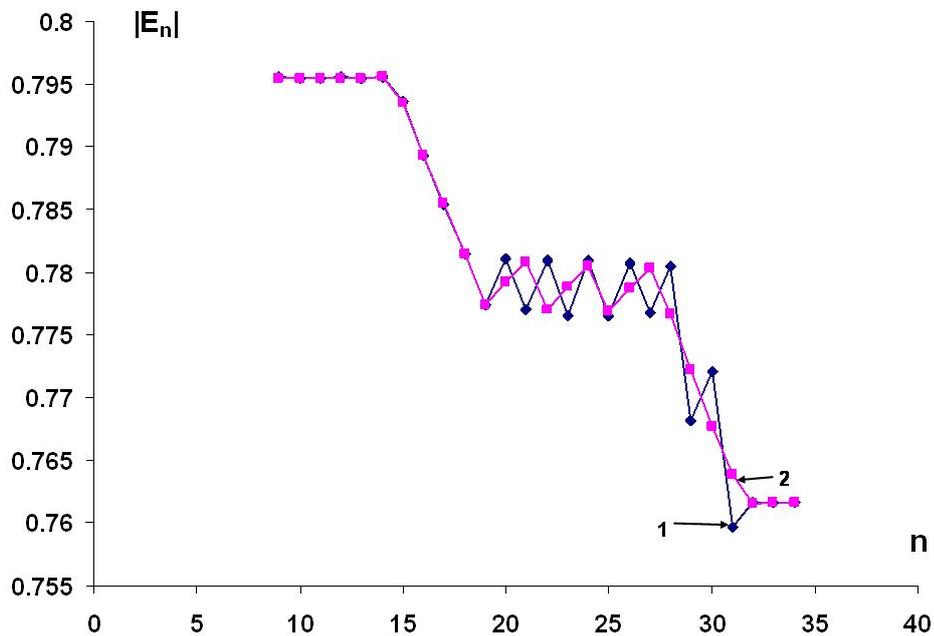

Fig. 10 Absolute values of longitudinal electric field on the middle of the cells for NM (1) and TCM (2) tuning methods

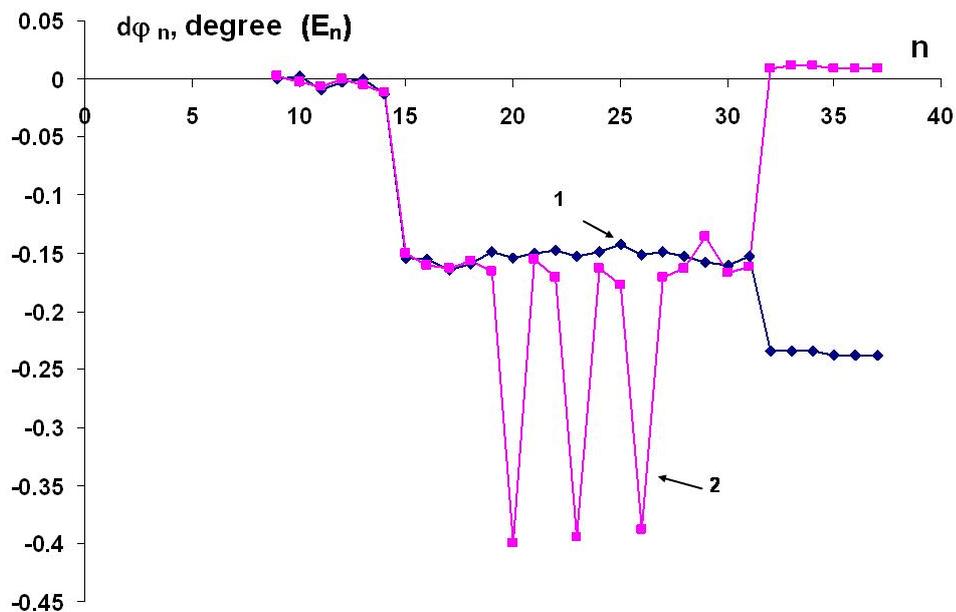

Fig. 11 Deviations of phase shift of longitudinal electric field in the middle of the cells from $2\pi n/3$ for NM (1) and TCM (2) tuning methods

### 3.2 SICA and KUT type DLW

The accelerating structures with strong current load have the greater parameter gradients and the increasing field distribution without current. Indeed, for the SLAC structure such value $(v_{group}^{(start)} - v_{group}^{(end)})/(v_{group}^{(start)} M)$, where M- number of cells, equals 0.008, while for the SICA [27] and KUT[28] structures it equals 0.0018. In this subsection we present the results of tuning fragment of DLW with radius distribution that is presented in Fig. 12 ($d$ =2.9148 cm, $t$ =0.5842 cm, $r$ =.0 cm).. The gradient of aperture radius is three times more than for the SLAC type DLW.

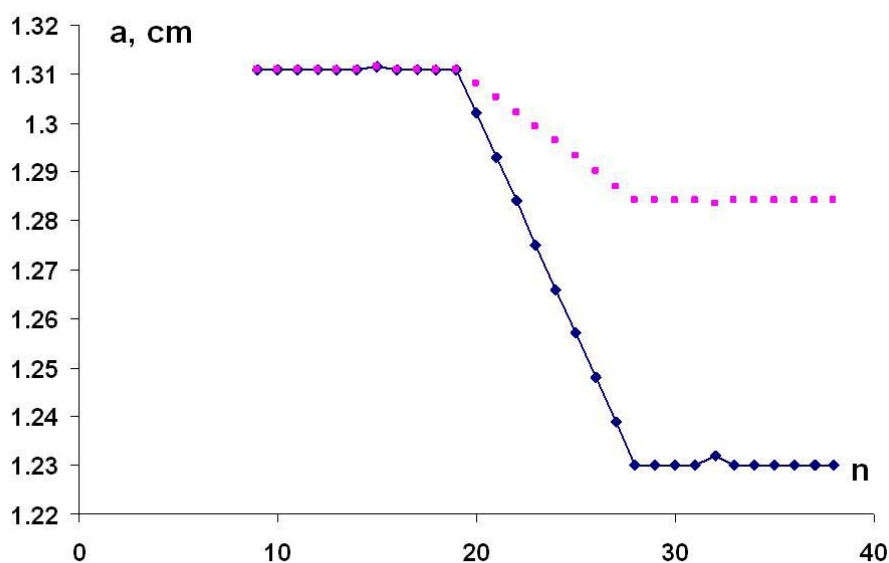

Fig. 12 Radii of cavities as a function of the cell number after TM tuning (blue). Red points – the case that was considered in subsection 3.1

The NM and TCM tuning procedures were the same as in subsection 3.1. But under TCM tuning procedure the reflection coefficient was achieved $R$ =1.8E-2. .

Differences of values of $G_n^{(3)}$ and $F_n^{(m)}$ (Fig. 13, Fig. 14) for NM and TCM tuning methods are approximately three times more than for the SLAC type DLW. Imaginary parts of $F_n^{(m)}$ for NM (1) and TCM (2) tuning methods (Fig. 15) are larger than for the SLAC type DLW and are approximately the constant values. This proves assertion that the imaginary part of $F_n^{(m)}$ is proportional to the field gradient (compare Fig. 9 and Fig. 16 ).

Characteristics of electromagnetic fields in the structure after tuning are presented in Fig. 16 and Fig. 17. From these results it follows that the DLW also have the similar electromagnetic field distributions under tuning by two different methods. But the phase deviation is greater than for the SLAC type DLW  and reach values of several degrees.

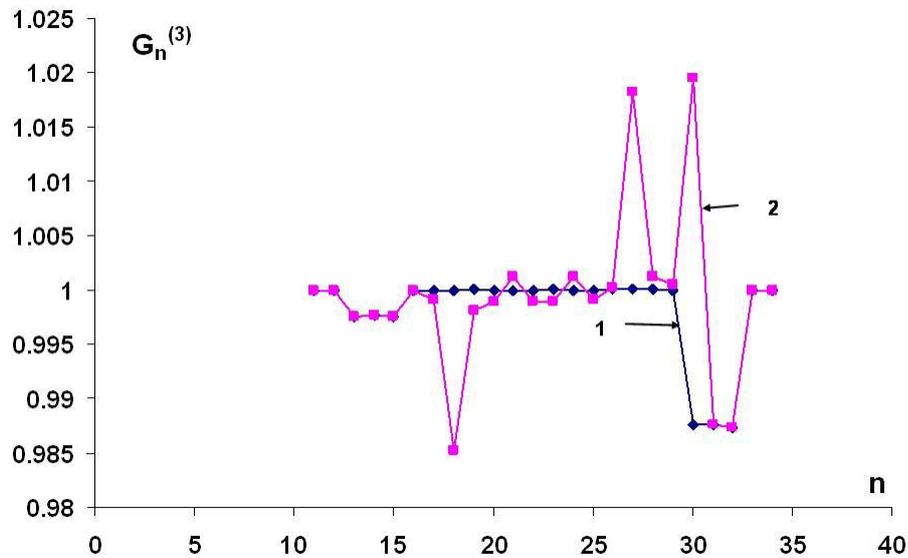

Fig. 13 $G_n^{(3)}$ for NM (1) and TCM (2) tuning methods

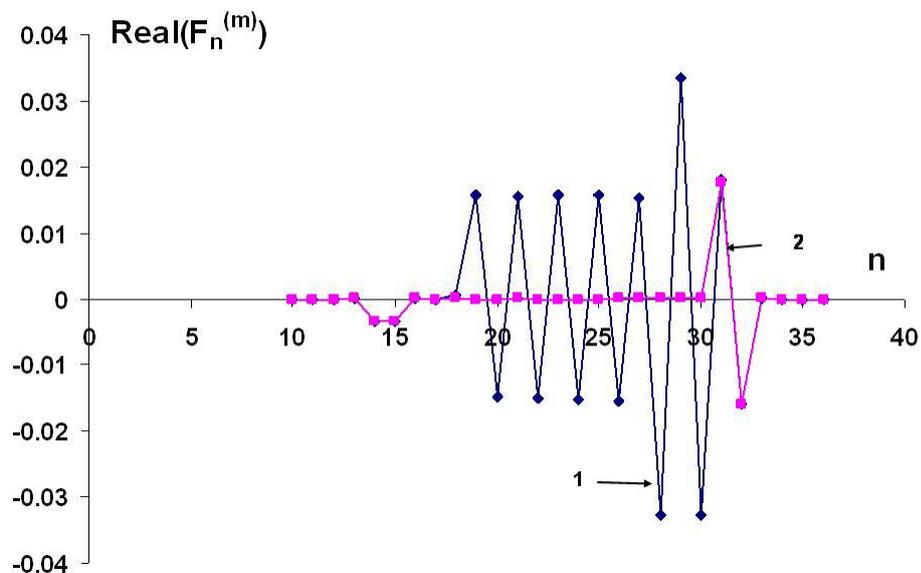

Fig. 14 Real part of $F_n^{(m)}$ for NM (1) and TCM (2) tuning methods

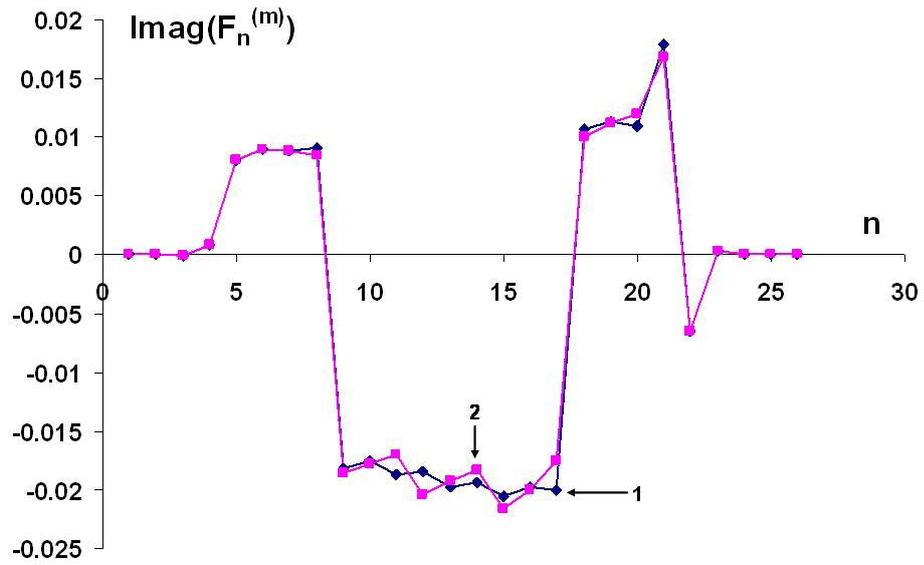

Fig. 15 Imaginary part of $F_n^{(m)}$ for NM (1) and TCM (2) tuning methods

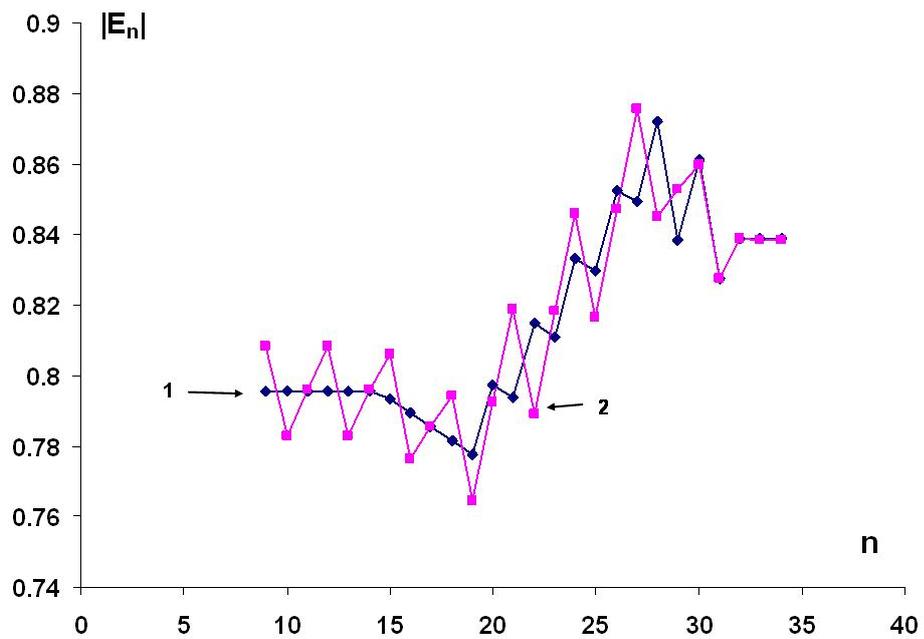

Fig. 16 Absolute values of longitudinal electric field on the middle of the cells for NM (1) and TCM (2) tuning methods

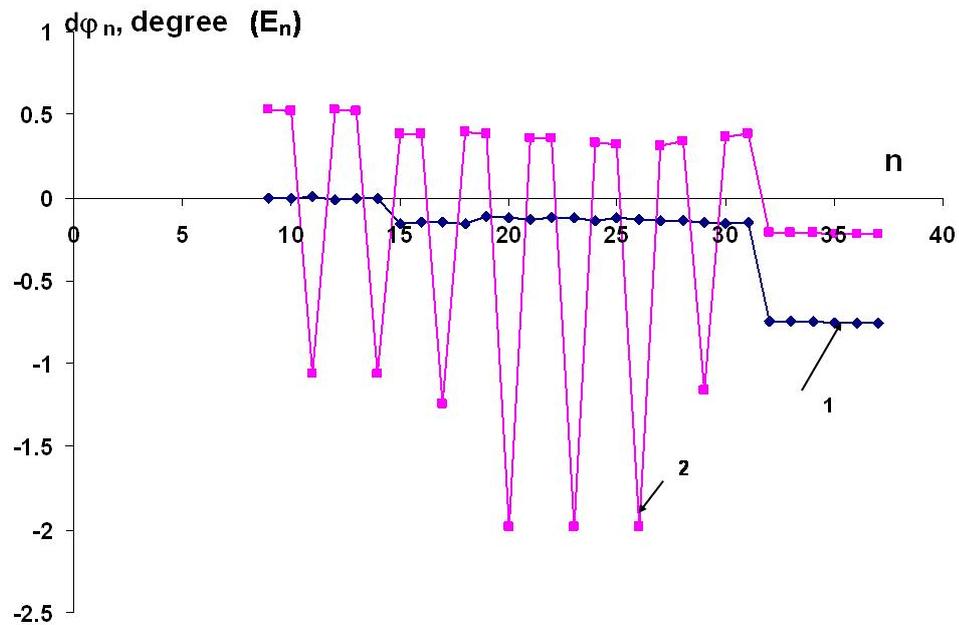

Fig. 17 Deviations of phase shift of longitudinal electric field in the middle of the cells from $2\pi n/3$ for NM (1) and TCM (2) tuning methods

### 3.2 Injector section

Injector sections are usually the DLW with geometrical sizes that provide changing along structure not only phase velocity, but the amplitude of accelerating field too. We studied the quality of the NM and TCM tuning methods for such DLWs. We chose geometrical sizes close[8] to the parameters of the injector that described in [29]. Dependences of the radii of apertures and the cavity length on the cell number are presented in Fig. 18 ($t$ =0.4 cm, $r$ =.0 cm, quality factors for cells N14÷27 were calculated by using SUPERFISF $Q$ = 9500÷1400).

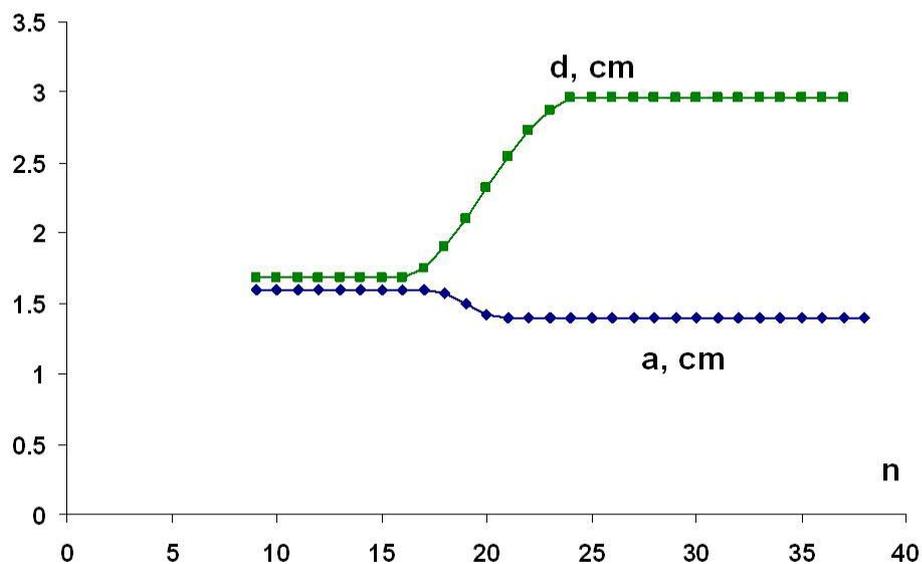

Fig. 18

─────────────

[8] We did not take into account roundings of the apertures.

The NM tuning procedure were the same as in subsection 3.1 and 3.2 and the reflection coefficient was achieved $R$ =4.E-4. But using TCM tuning procedure was not such simple as in subsection 3.1 and 3.2, especially for cells N17÷21, where the radii of apertures change. The main problem is the dependence of $F_n^{(m)}$ on the parameters of the adjacent cells. When we tune the certain cell ($n = s$) values of $F_{s\pm1}^{(m)}$ (and even $F_{s\mp2}^{(m)}$) change. So, we have to make several steps forward and backward to find the minimum values of $F_n^{(m)}$. After tuning the reflection coefficient was achieved $R$ =3.8E-2.

Results of tunings are presented in Fig. 19 - Fig. 23. The most important result is that deviations of phase shifts of longitudinal electric field in the middle of the cells from $2\pi n/3$ for TCM (2) tuning method has nonperiodic character and can not be made less than 5 degree. Therefore, the total particle phase shift relative field will not be small. It is difficult to evaluate the influence of such phase distribution on particle dynamic without conducting simulation of particle dynamics.

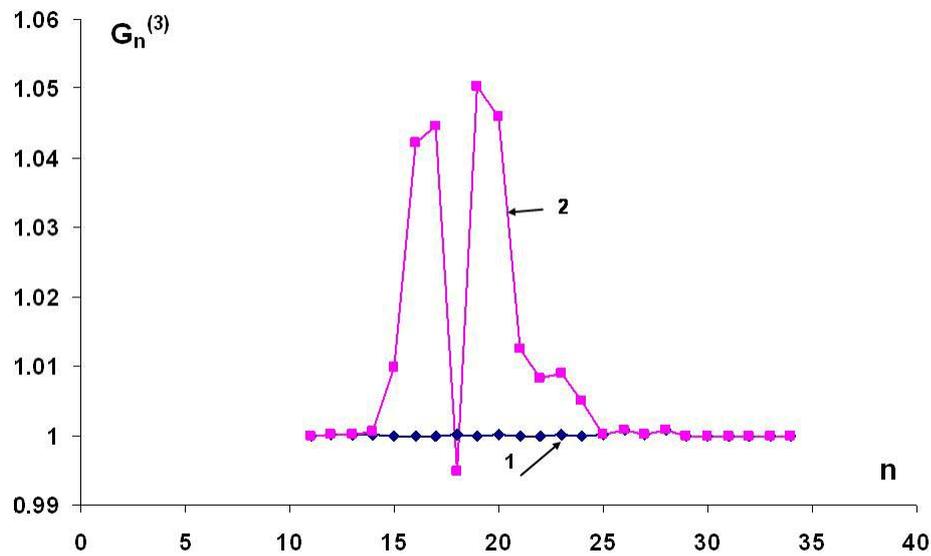

Fig. 19 $G_n^{(3)}$ for NM (1) and TCM (2) tuning methods

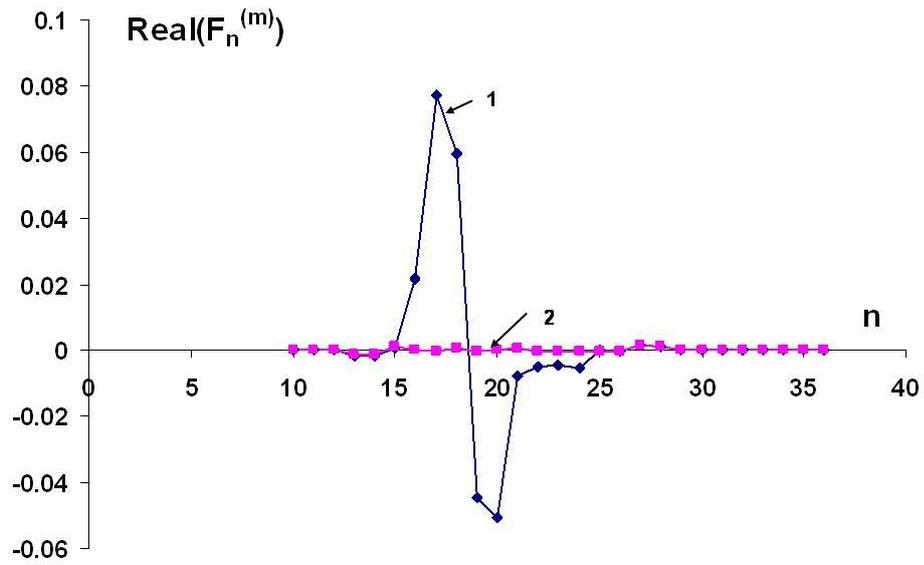

Fig. 20 Real part of $F_n^{(m)}$ for NM (1) and TCM (2) tuning methods

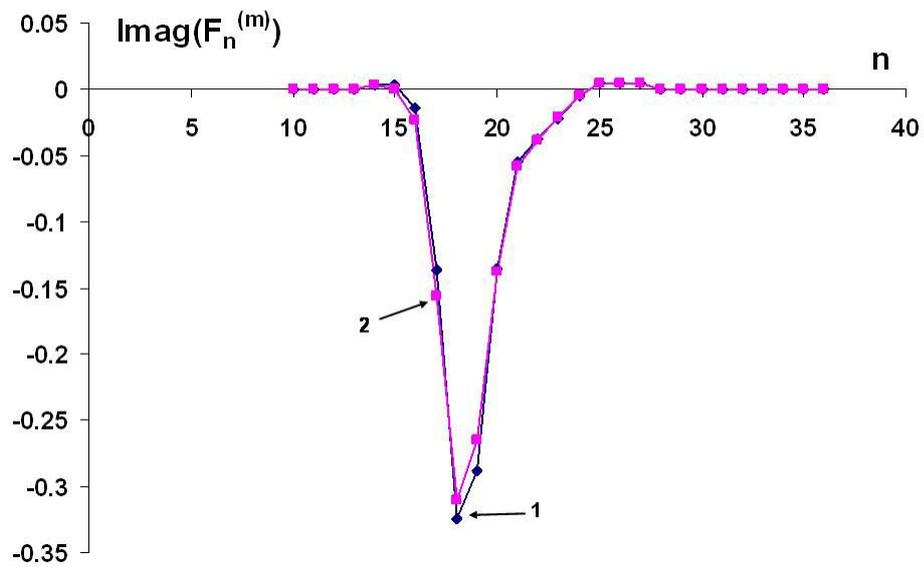

Fig. 21 Imaginary part of $F_n^{(m)}$ for NM (1) and TCM (2) tuning methods

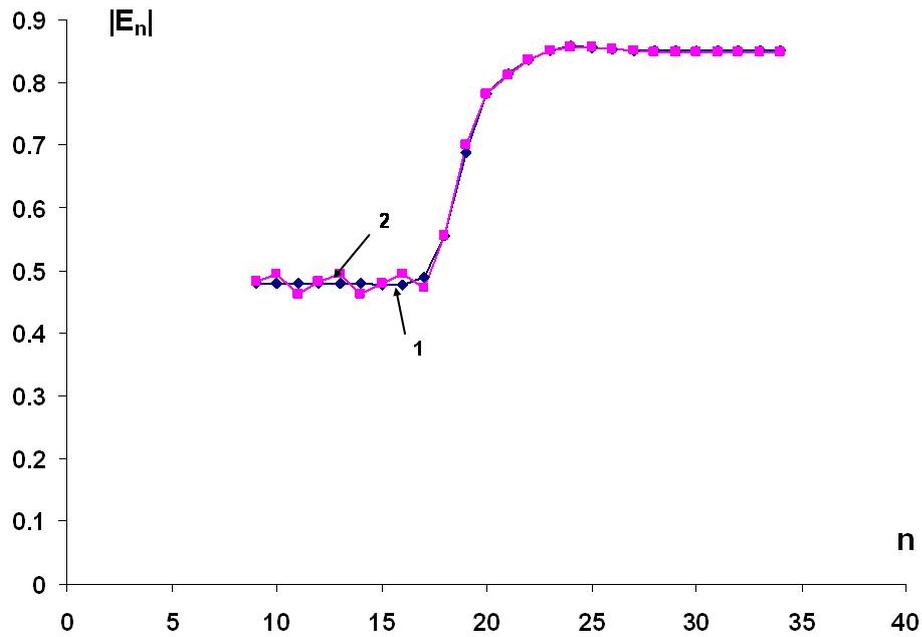

Fig. 22 Absolute values of longitudinal electric field on the middle of the cells for NM (1) and TCM (2) tuning methods

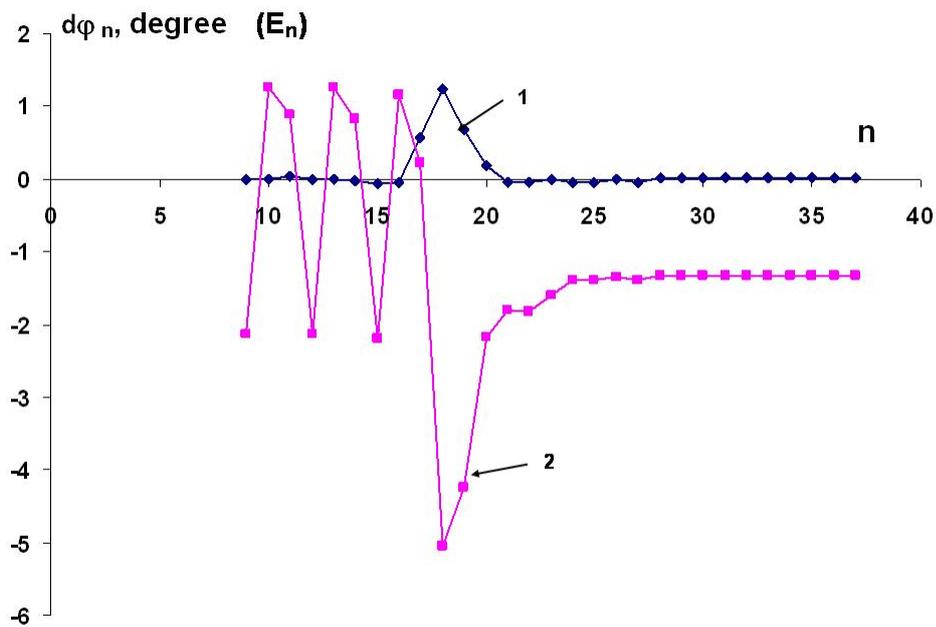

Fig. 23 Deviations of phase shifts of longitudinal electric field in the middle of the cells from $2\pi n/3$ for NM (1) and TCM (2) tuning methods

## Conclusions

We obtain some results that can be useful in the process of post-tuning of nonunifrom DLWs. On the base of more general model we assessed accuracy of method that is widely used for post- tuning accelerator sections.

Our consideration has shown that for wide range of DLW parameters using simple "tuning" coefficients gives appropriate characteristics of DLWs. For injector section, in which phase velocity and amplitude change, there are some deviation from the desired characteristics. It is necessary to conduct additional simulation of particle dynamics to evaluate the influence of obtained phase distribution on beam characteristics.

Appendix 1

We now consider the general problem obtaining the coupling cavity equations for arbitrary waveguides.

Let's divide the waveguide[9] into the set of arbitrary volumes $V_k$ and represent electromagnetic field as

$$\vec{E} = \tilde{\vec{E}} + \begin{cases} \sum_q A_q^{(k)} \vec{E}_q^{(k)}, \ \vec{r} \in V_k, \\ 0, \qquad\qquad \vec{r} \notin V_k \end{cases} \tag{1.9}$$

$$\vec{H} = \tilde{\vec{H}} + \begin{cases} \sum_q A_q^{(k)} \dfrac{\omega}{\omega_q^{(k)}} \vec{H}_q^{(k)}, \ \vec{r} \in V_k, \\ 0, \qquad\qquad\qquad \vec{r} \notin V_k \end{cases} \tag{1.10}$$

where $\vec{E}_q^{(k)}, \vec{H}_q^{(k)}$ - field distribution of some eigen oscillations (that we will consider as the basic oscillations) in the volume $V_k$, $\omega_q^{(k)}$ - eigen frequency. Vectors $\vec{E}_q^{(k)}, \vec{H}_q^{(k)}$ obey such equations

$$\begin{aligned} rot \ \vec{E}_q^{(k)} &= i \, \omega_q^{(k)} \mu_0 \vec{H}_q^{(k)} \ , \\ rot \ \vec{H}_q^{(k)} &= -i \, \omega_q^{(k)} \varepsilon_0 \vec{E}_q^{(k)}, \ \ \vec{r} \in V_k \end{aligned} \tag{1.11}$$

with boundary conditions $E_{q,\tau}^{(k)} = 0$ on the volume borders.

Substituting (1.9) into Maxwell equation we obtain

$$rot \ \tilde{\vec{E}} - i \, \omega \mu_0 \tilde{\vec{H}} = \begin{cases} \sum_q i \mu_0 \dfrac{\omega^2 - \omega_q^{(k)2}}{\omega_q^{(k)}} A_q^{(k)} \vec{H}_q^{(k)}, \ \vec{r} \in V_k, \\ 0, \qquad\qquad\qquad\qquad \vec{r} \notin V_k \end{cases} , \tag{1.12}$$

$$rot \tilde{\vec{H}} + i \, \omega \varepsilon_0 \tilde{\vec{E}} = 0$$

We can represent $\tilde{\vec{E}}, \tilde{\vec{H}}$ as

$$\begin{aligned} \tilde{\vec{E}} &= \sum_q \sum_{k=-\infty}^{k=\infty} A_q^{(k)} \tilde{\vec{E}}_q^{(k)}, \\ \tilde{\vec{H}} &= \sum_q \sum_{k=-\infty}^{k=\infty} A_q^{(k)} \tilde{\vec{H}}_q^{(k)} \end{aligned} \tag{1.13}$$

where $\tilde{\vec{E}}_q^{(k)}, \tilde{\vec{H}}_q^{(k)}$ are the solutions of such equations

$$rot \ \tilde{\vec{E}}_q^{(k)} - i \, \omega \mu_0 \tilde{\vec{H}}_q^{(k)} = i \mu_0 \dfrac{\omega^2 - \omega_q^{(k)2}}{\omega_q^{(k)}} \vec{H}_q^{(k)} \eta(\vec{r} - V_k), \tag{1.14}$$

$$rot \ \tilde{\vec{H}}_q^{(k)} + i \, \omega \varepsilon_0 \tilde{\vec{E}}_q^{(k)} = 0,$$

where $\eta(\vec{r} - V_k) = \begin{cases} 1, \vec{r} \in V_k \\ 0, \vec{r} \notin V_k \end{cases}$ .

---

[9] The waveguide can be infinitive or restricted by metal plates

Equations for $A_q^{(n)}$ have such form

$$\left(\omega_q^{(n)2} - \omega^2\right) A_q^{(n)} = \frac{i\,\omega_q^{(n)}}{N_q^{(n)}} \left(\oint_{S_n} [\vec{E}_\tau^{(n)} \vec{H}_q^{(n)*}] d\vec{S} + \oint_{S_{n+1}} [\vec{E}_\tau^{(n+1)} \vec{H}_q^{(n)*}] d\vec{S}\right) =$$

$$= \frac{i\,\omega_q^{(n)}}{N_q^{(n)}} \sum_{q'} \sum_{k=-\infty}^{k=\infty} A_{q'}^{(k)} \left(\oint_{S_n} [\vec{\tilde{E}}_{q',\tau}^{(k)} \vec{H}_q^{(n)*}] d\vec{S} + \oint_{S_{n+1}} [\vec{\tilde{E}}_{q',\tau}^{(k)} \vec{H}_q^{(n)*}] d\vec{S}\right) = \sum_{q'} \sum_{k=-\infty}^{k=\infty} A_{q'}^{(k)} \alpha_{q,q'}^{(n,k)}$$

(1.15)

where $S_n$ and $S_{n+1}$ are the surfaces through which the $V_n$ volume contacts with the outside part of waveguide.

We obtained coupling cavity equations in the general form. If we find $\vec{H}_q^{(n)}, \omega_q^{(n)}, \vec{\tilde{E}}_q^{(k)}$, we can calculate the coupling coefficients $\alpha_{q,q'}^{(n,k)}$.

### References

1 Charles W.Steele A nonresonant perturbation theory.. IEEE Trans Microwave Theory Tech, 1966, MTT-14, N 2, pp.70-74

2 K. B. Mallory ; R. H. Miller On Nonresonant Perturbation Measurements IEEE Trans Microwave Theory Tech, 1966, MTT-14, N 2, pp.99-100

3 S.M. Hanna, G.B. Bowden, H.A. Hoag et al. A semi-automated system for the characterization of NLC accelerating structures. Proceedings of PAC95, pp.1008-1110

4 S. M. Hanna, G. B. Bowden, H. A. Hoag et all. Microwave Cold-Testing Techniques for the NLC. Proceedings of EPAC96

5 S.M. Hanna, R.J. Loewen, H.A. Hoag Development of characterization techniques for x-band accelerator structures. Proceedings of PAC97, pp.539-541

6 H.Carter, D.Finley, I.Gonin et al. Automated microwave low power testing techniques for NLC. Proceedings of LINAC2002, pp.761-763

7 A.Mostacci, L.Palumbo, R.Da.Re et al. About non resonant perturbation field measurement in standing wave cavities. Proceedings of PAC09, pp.3407-3409

8 T.Khabiboulline, V.Puntus, M.Dohlus et al. A new tuning method for traveling wave structures. Proceedings of PAC95, pp.1666-1668

9 T.Khabiboulline, M.Dohlus, N.Holtkamp. Tuning of a 50-cell costant gradient S-band travelling wave accelerating structure by using a nonresonant perturbation method. Internal Report DESY M-95-02, 1995

10 T. Khabiboulline, T. Arkan, E. Borissov et al. Development of X-band accelerating structures at FERMILAB. Proceedings of PAC03, pp.1210-1212

11 Jiaru Shi, Alexej Grudiev, Andrey Olyunin Tuning of CLIC accelerating structure prototypes at CERN. Proceedings of LINAC2010, pp.97-99

12 W. C. Fang, Z. T. Zhao et al. R&D of C-band accelerating structure at SINAP. Proceedings of LINAC2010, pp.199-201

13 Fang Wen Cheng, Tong De Chun, Gu Qiang et al. Design and experimental study of a C-band traveling-wave accelerating structure. Chinese Science Bulletin, 2011, v.56, N.1, pp.18-23

14 W. C. Fang, Q. Gu, Z. T. Zhao, D. C. Tong. The nonresonant perturbation theory based field measurement and tuning of a linac accelerating structure. Proceedings of LINAC2012, pp.375-377

15 Ang Wen Cheng, Tong De Chun, Gu Qiang et al.The nonresonant perturbation theory based field measurement and tuning of a linac accelerating structure. Science China, 2013, Vol.56, N.11, pp.2104–2109 16 Gong Cun-Kui. Zheng Shu-Xin, Shao Jia-Hang et al. A tuning method for nonuniform traveling-wave accelerating structures. Chinese Physics C, 2013, V.37, N.1, 017003

17 Jiaru Shi, Alexej Grudiev, WalterWuensch Tuning of X-band traveling-wave accelerating structures Nuclear Instruments and Methods in Physics Research 2014, A704, pp.14-18

18 Y.C.Nien, C.Liebig, M.Hüning et al. Tuning of 2.998 GHz S-band hybrid buncher for injector upgrade of LINAC II at DESY. Nuclear Instruments and Methods in Physics Research 2014, A761, pp.69–78

19 R. Wegner, W. Wuensch, Bead-pull measurement method and tuning of a prototype CLIC crab cavity, Proceedings of LINAC2014, pp.134-136

20 Jianhao Tan, Dechun Tong, Qiang Gu et al. Development of non-resonant perturbing method for tuning traveling wave deflecting structures. Proceedings of IPAC2015, pp.2985-2987

21 G. Waldschmidt, T. Smith, L. Morrison et al. Transverse cavity tuning at the advanced photon source. Proceedings of IPAC2016

22 M.I. Ayzatskiy, V.V. Mytrochenko. Electromagnetic fields in nonuniform disk-loaded waveguides. http://lanl.arxiv.org/ftp/arxiv/papers/1503/1503.05006.pdf, LANL.arXiv.org e-print archives, 2015

23 M.I. Ayzatskiy, V.V. Mytrochenko. Coupled cavity model based on the mode matching technique. http://lanl.arxiv.org/ftp/arxiv/papers/1505/1505.03223.pdf, LANL.arXiv.org e-print archives, 2015

24 M.I. Ayzatskiy, V.V. Mytrochenko. Coupled cavity model for disc-loaded waveguides. http://lanl.arxiv.org/ftp/arxiv/papers/1511/1511.03093.pdf, LANL.arXiv.org e-print archives, 2015

25 M. I. Ayzatsky, V. V. Mytrochenko. Numerical design of nonuniform disk-loaded waveguides. http://lanl.arxiv.org/ftp/arxiv/papers/1604/ 1604.05511.pdf, LANL.arXiv.org e-print archives, 2016

26 R.B Neal, General Editor, The Stanford Two-Mile Accelerator, New York, W.A. Benjamin, 1968

27 E. Jensen. CTF3 drive beam accelerating structures Proceedings of LINAC2002, pp.34-36

28 M.I.Ayzatsky, E.Z.Biller. Development of Inhomogeneous Disk-Loaded Accelerating waveguides and RF-coupling. Proceedings of Linac96, p.119-121

29 M.I.Ayzatsky, , A.N.Dovbnya, V.F.Zhiglo et al. Accelerating System for an Industrial linac. Problems of Atomic Science and Technology. Series "Nuclear Physics Investigations". 2012, No4, p.24-28